\documentstyle[11pt,moriond]{article}
\input epsf.tex

\def\Journal#1#2#3#4{{#1} {\bf #2}, #3 (#4)}
\def\PRL{\em Phys. Rev. Lett.}
\def\PRD{{\em Phys. Rev.} D}

\begin{document}

\centerline{\hfill\vbox to 25 pt
{\hbox{CPT-99/P.3802}\hbox{gr-qc/9903058}\vfil}}
\vspace*{-1.5cm}

\vspace*{4cm}
\title{COMPARING SOLAR-SYSTEM, BINARY-PULSAR,\\
AND GRAVITATIONAL-WAVE TESTS OF GRAVITY\,\footnote{Invited talk at
the XXXIVth Rencontres de Moriond, Les Arcs
(France), January 23--30 1999.}}

\author{Gilles ESPOSITO-FARESE}

\address{Centre de Physique Th\'eorique, CNRS Luminy, Case 907,\\
F 13288 Marseille Cedex 9, France}

\maketitle\abstracts{
This talk is based on my work in collaboration with Thibault Damour.
We compare the probing power of different classes of gravity
experiments: solar-system tests (weak-field regime),
binary-pulsar tests (strong-field regime), and future
gravitational-wave observations of inspiralling binaries
(strong-field effects detected in our weak-gravitational-field
conditions). This is done within the most natural class of
alternative theories to general relativity, namely tensor-scalar
theories, in which the gravitational interaction is mediated by one
tensor field ($g_{\mu\nu}$) together with one or several scalar
fields ($\varphi$). Our main conclusion is that strong-field tests
are qualitatively different from weak-field experiments: They
constrain theories which are strictly indistinguishable from general
relativity in the solar system. We also show that binary-pulsar data
are so precise that they already rule out the theories for which
scalar effects could have been detected with {\sc LIGO} or {\sc
VIRGO}. This proves that it is therefore sufficient to compute the
`chirp' templates within general relativity.}

\section{Introduction}

Solar-system experiments probe the weak-field regime of gravity.
Indeed, the largest deviations from the flat metric are at the
surface of the Sun, and are of order $Gm_\odot/R_\odot c^2
\approx 2 \times 10^{-6}$ (where $m_\odot$ and $R_\odot$ denote the
mass and the radius of the Sun). On the other hand, binary-pulsar
data allow us to test the strong-field regime, since the compactness
$Gm/Rc^2$ of a neutron star is of order $0.2$ (not far from the
theoretical maximum of $0.5$, corresponding to black holes).
The forthcoming gravitational-wave observations of inspiralling
compact binaries, by interferometers like {\sc LIGO} or {\sc VIRGO},
will also allow us to test gravity in strong-field conditions. It is
therefore interesting to compare and contrast the probing power of
these different classes of experiments.

\begin{figure}
\centerline{\epsfbox{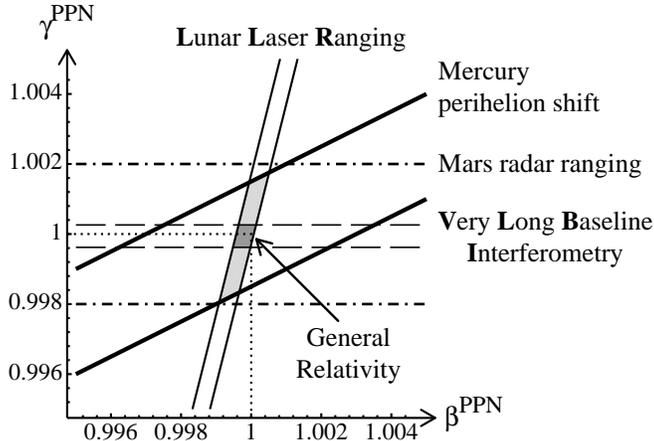}}
\caption{Solar-system constraints on the Eddington parameters. The
allowed region is shaded.
\label{fig:1}}
\end{figure}

A convenient quantitative way of doing this comparison is to embed
general relativity into a class of alternative theories. For
instance, the Parametrized Post-Newtonian (PPN) formalism is
very useful to study weak-field gravity, at order $1/c^2$ with
respect to the Newtonian interaction. The original idea was
formulated by Eddington,\cite{edd} who wrote the usual
Schwarzschild metric in isotropic coordinates, but introduced some
phenomenological parameters $\beta^{\rm PPN}$, $\gamma^{\rm PPN}$,
in front of the different powers of the dimensionless ratio
$Gm/rc^2$~:
\begin{equation}
-g_{00} = 1-2{Gm\over rc^2}+2\beta^{\rm
PPN}\left({Gm\over rc^2}\right)^2+O\left({1\over
c^6}\right),\qquad
g_{ij} = \delta_{ij}\left[1+2\gamma^{\rm PPN}{Gm\over
rc^2}+O\left({1\over c^4}\right)\right].
\label{eq:1}
\end{equation}
General relativity, which corresponds to $\beta^{\rm PPN} =
\gamma^{\rm PPN} = 1$, is thus embedded into a two-dimensional space
of theories. [The third parameter that one may introduce in front of
$Gm/rc^2$ in $g_{00}$ can be reabsorbed in the definition of the
mass $m$.] The constraints imposed in this space by solar-system
experiments are displayed in Fig.~\ref{fig:1}. We have also indicated
the tight bounds obtained in 1997 with Very Long Baseline
Interferometry (VLBI), although they are not yet published.\cite{vlbi}
See K.~Nordtvedt's contribution to the present Proceedings for even
more precise limits recently extracted from Lunar Laser Ranging (LLR)
data. For the most general formulation of the PPN formalism, see the
works of Nordtvedt \& Will.\cite{wn,willbook}

Figure~\ref{fig:1} shows clearly that the different
solar-system experiments are complementary. We wish to compare
the strong-field tests in a similar way, but an extension of the PPN
formalism to all orders in $1/c^n$ would need the introduction of an
infinite number of phenomenological parameters. It will be more
convenient to focus instead on the most natural class of alternative
theories to general relativity.

\section{Tensor-scalar theories of gravity}

The existence of scalar partners to the graviton is predicted by all
unified and extra-dimensional theories, notably superstrings.
Moreover, tensor-scalar theories are the only consistent massless
field theories able to satisfy exactly the weak equivalence
principle (universality of free fall of laboratory-size objects).
They are also the only known theories satisfying ``extended Lorentz
invariance'', {\it i.e.}, such that the physics of
subsystems, influenced by external masses, exhibit Lorentz
invariance. Finally, they explain the key role played by $\beta^{\rm
PPN}$ and $\gamma^{\rm PPN}$ in the PPN formalism (all the 8 extra
parameters introduced by Will and Nordtvedt\,\cite{wn,willbook} vanish
identically), and they are general enough to describe many
possible deviations from general relativity. These reasons show
that tensor-scalar theories are privileged alternatives to
general relativity.

Like in Einstein's theory, the action of matter is given by a
functional $S_m[\psi_m,\widetilde g_{\mu\nu}]$ of some matter fields
$\psi_m$ (including gauge bosons) and one second-rank symmetric
tensor\,\footnote{To simplify, we will consider here only theories
which satisfy exactly the weak equivalence principle.} $\widetilde
g_{\mu\nu}$. The difference with general relativity lies in the
kinetic term of $\widetilde g_{\mu\nu}$. Instead of being a pure
spin-2 field, it is here a mixing of spin-2 and spin-0 excitations.
More precisely, it can be written as $\widetilde g_{\mu\nu} = {\rm
exp}[2a(\varphi)]g_{\mu\nu}$, where $a(\varphi)$ is a function of a
scalar field\,\footnote{To simplify again, we will restrict our
discussion to a single scalar field, although the case of
tensor-multi-scalar theories can also be treated in a
similar way.\cite{def1}} $\varphi$, and $g_{\mu\nu}$ is the Einstein
(spin-2) metric. The action of the theory reads thus
\begin{equation}
S = {c^3\over 16\pi
G}\int{d^4x\sqrt{-g}\Bigl(R-2g^{\mu\nu}\partial_\mu\varphi
\partial_\nu\varphi\Bigr)}
+S_m\left[\psi_m,e^{2a(\varphi)}g_{\mu\nu}\right].
\label{eq:2}
\end{equation}
[Our signature is $\scriptstyle - + + +$, $R$ is the scalar curvature
of $g_{\mu\nu}$, and $g$ its determinant.]

\begin{figure}
\centerline{\epsfbox{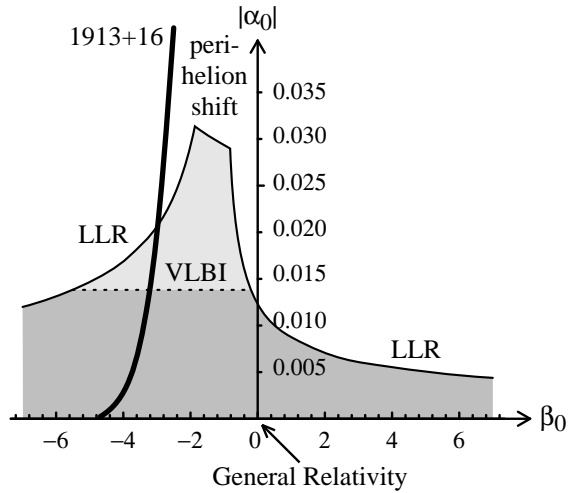}}
\caption{Solar-system constraints on tensor-scalar theories. The
allowed region is shaded. The region consistent with PSR 1913+16
data lies on the right of the bold line.
\label{fig:2}}
\end{figure}

Our discussion will now be focused on the function $a(\varphi)$,
which characterizes the coupling of matter to the scalar field. It
will be convenient to expand it around the background value
$\varphi_0$ of the scalar field ({\it i.e.}, its value far from any
massive body):
\begin{equation}
a(\varphi) =
\alpha_0(\varphi-\varphi_0)
+{1\over2}\beta_0(\varphi-\varphi_0)^2
+ {1\over 3!}\beta_0'(\varphi-\varphi_0)^3
+\cdots\ ,
\label{eq:3}
\end{equation}
where $\alpha_0$, $\beta_0$, $\beta_0'$, \dots\ are constants defining
the theory. In particular, the slope $\alpha_0$ measures the coupling
strength of the linear interaction between matter and the scalar
field. This slope and the curvature $\beta_0$ of $a(\varphi)$ are
the only parameters which appear at the post-Newtonian order, {\it
i.e.}, when measuring effects of order $1/c^2$ in weak-field
conditions. The same solar-system experiments as those of
Fig.~\ref{fig:1} give now the constraints displayed in
Fig.~\ref{fig:2}.
[The bold line will be discussed in the next section.] We see that
the main information provided by weak-field tests is that the slope
$\alpha_0$ must be small, {\it i.e.}, that the background value
$\varphi_0$ must correspond almost to an extremum of $a(\varphi)$. On
the other hand, the curvature $\beta_0$ is not constrained at all if
$\alpha_0$ is small enough. In particular, its sign is not known, and
$\varphi_0$ may thus be close either to a maximum ($\beta_0<0$) or to
a minimum ($\beta_0>0$) of the function $a(\varphi)$. [Note that
action (\ref{eq:2}) contains only positive-energy excitations,
independently of the shape of the function $a(\varphi)$, and
therefore independently of the sign of $\beta_0$.]

Let us underline that in Fig.~\ref{fig:2}, the vertical axis
($\beta_0 = 0$) corresponds to the Jordan-Fierz-Brans-Dicke theory,
where the constant slope $\alpha_0$ is related to the usual
Brans-Dicke parameter by $\alpha_0^2 = 1/(2\omega_{\rm BD} + 3)$.
On the other hand, the horizontal axis ($\alpha_0 = 0$) corresponds
to theories which are {\it perturbatively equivalent\/} to general
relativity. Indeed, it can be shown\,\cite{def4} that any deviation
{}from Einstein's theory (at any order $1/c^n$) involves a factor
$\alpha_0^2$.

However, {\it nonperturbative\/} effects can occur in strong-field
conditions,\cite{def3} even if $\alpha_0$ vanishes. Indeed, the
presence of matter induces an effective potential for the scalar
field. If the curvature parameter $\beta_0$ is negative and the
compactness $Gm/Rc^2$ of a body is large enough, one finds that this
effective potential has the shape of a Mexican hat, with a local
maximum at $\varphi = \varphi_0$ and two symmetric minima at
nonvanishing values of $\varphi - \varphi_0 = \pm \varphi_{\rm min}$.
It is therefore energetically favorable for the body to create a
scalar field which differs from the background $\varphi_0$. This
phenomenon is similar to the spontaneous magnetization of
ferromagnets, and we have called it ``spontaneous
scalarization''.\cite{def7} The theory now behaves like in weak-field
conditions, but with a matter-scalar coupling strength proportional
to $a'(\varphi_0\pm\varphi_{\rm min}) \approx
\pm\beta_0 \varphi_{\rm min} \neq 0$ instead of $\alpha_0 \approx 0$.
Large deviations from general relativity can thus occur in systems
involving compact bodies, like neutron stars, even if the theory is
indistinguishable from general relativity in the solar system. These
conclusions have been confirmed by numerical integrations of the
coupled differential equations of the metric $g_{\mu\nu}$ and the
scalar field $\varphi$, while using either
polytropes\,\cite{def3,def7} or realistic equations of
state\,\cite{def9} to describe nuclear matter inside a neutron star.

\section{Binary-pulsar tests}

A pulsar is a rapidly rotating neutron star emitting radio waves in
a particular direction, like a lighthouse. Experiment tells us that
isolated pulsars are very stable clocks. A binary pulsar (a pulsar
and a companion orbiting around each other) is thus a {\it moving
clock}, the best tool that one could dream of to test a relativistic
theory. Indeed, the Doppler effect modifies the frequency of the
pulses, and the time between two maxima of this frequency is thus a
measure of the orbital period $P_b$. A careful analysis\,\cite{dt} of
the Times Of Arrivals can give in fact many other orbital data,
like the eccentricity $e$, the angular position of the periastron
$\omega$, as well as the measure of several relativistic effects.

In the case of PSR 1913+16, three post-Keplerian parameters are
determined with great accuracy:\,\cite{t} (i)~an observable denoted
$\gamma_{\rm Timing}$, which combines the second-order Doppler effect
and the redshift due to the companion; (ii)~the periastron advance
$\dot\omega$; (iii)~the variation of the orbital period, $\dot P_b$,
due to gravitational radiation reaction. For a given theory of
gravity, these {\bf 3} quantities can be predicted in terms of the
{\bf 2} unknown masses $m_A$, $m_B$, of the pulsar and its companion.
This gives (3$-$2=) {\bf 1} test of the theory: There must exist a
pair of masses $(m_A,m_B)$ such that the three equations $\gamma_{\rm
Timing}^{\rm th}(m_A,m_B) = \gamma_{\rm Timing}^{\rm obs}$,
$\dot\omega^{\rm th}(m_A,m_B) = \dot\omega^{\rm obs}$, $\dot P_b^{\rm
th}(m_A,m_B) = \dot P_b^{\rm obs}$, are simultaneously satisfied. It
is shown in J.~Taylor's contribution to the present Proceedings that
general relativity passes this test with flying colors: In the plane
of the masses $(m_A,m_B)$, the three curves defined by the above
equations meet in one point.

The predictions of tensor-scalar theories for these three
observables can be computed for any shape of the function
$a(\varphi)$. However, to make a quantitative comparison between this
binary-pulsar test and solar-system experiments, let us consider a
{\it generic\/} class of theories defined by parabolic functions
$a(\varphi)$, {\it i.e.}, such that the slope $\alpha_0$ and the
curvature $\beta_0$ are the only nonvanishing coefficients in
expansion (\ref{eq:3}). We find that all the theories lying on
the left of the bold line, in Fig.~\ref{fig:2}, are ruled out
by PSR 1913+16 data. Note in particular that the theories
corresponding to $\alpha_0 = 0$, $\beta_0 < -4.5$, are excluded by
this pulsar test\,\footnote{It is interesting to note that
binary-pulsar data are nicely consistent with the
results of tensor-scalar cosmological models,\cite{dn2} which
privilege the positive values of $\beta_0$.}, although they are
strictly indistinguishable from general relativity in the solar
system. This is due to the ``spontaneous scalarization'' of neutron
stars which occurs when $\beta_0$ is negative (see Sec.~2).

\begin{figure}
\centerline{\epsfbox{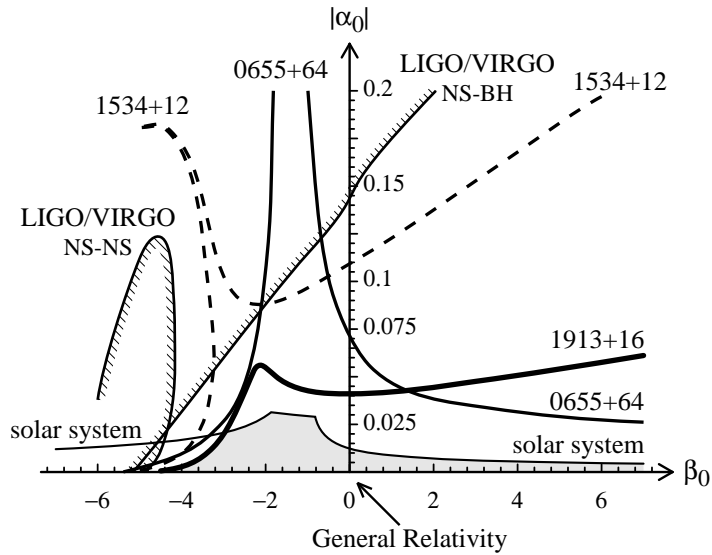}}
\caption{Combined experimental constraints on generic tensor-scalar
theories. The allowed region is shaded.
\label{fig:3}}
\end{figure}

The constraints imposed by two other binary-pulsar tests are
displayed in Fig.~\ref{fig:3}.
See I.~Stairs' contribution to the present Proceedings for a
discussion of PSR 1534+12. The region which lies on the right of
the dashed line is consistent with its four observables\,\cite{s}
$\gamma_{\rm Timing}$, $\dot\omega$, $r$, and $s$ (the last two
denoting the range and the shape of the Shapiro time delay).
PSR 0655+64 is a dissymmetrical system composed of a neutron star and
a white dwarf companion, which generically emits strong scalar
dipolar waves in tensor-scalar theories. The region consistent with
the small observed value of its $\dot P_b$ lies between the two solid
lines of Fig.~\ref{fig:3}.

\section{Detection of gravitational waves}

One of the main differences between general relativity
and tensor-scalar theories occurs in the expression of the energy
flux due to the emission of gravitational waves. It has the
form\,\cite{willbook,def1}
\begin{equation}
\left\{{{\rm Quadrupole}\over c^5}\right\}_{\rm spin\ 2}
\!\!+\left\{{{\rm Monopole}\over c}\left(\dot\sigma+{1\over
c^2}\right)^2+ {{\rm Dipole}\over c^3}+
{{\rm Quadrupole}\over c^5}\right\}_{\rm spin\ 0}
\!\!+ O\left({1\over c^7}\right),
\label{eq:4}
\end{equation}
where the first curly brackets contain the predictions of Einstein's
theory, and the second ones the extra contributions due to the
emission of scalar waves. The symbol $\sigma$ means the ``scalar
charge'' of the system, which is constant if the bodies are at
equilibrium, like in the above binary pulsars. The predictions for
their $\dot P_b$ are thus dominated by the dipolar term $\propto
1/c^3$.

On the other hand, in the case of a collapsing star, the time
derivative of this scalar charge may be of order unity, and one
expects thus a huge emission of monopolar waves $\propto 1/c$, much
larger than the general relativistic prediction $\propto 1/c^5$.
However, Fig.~\ref{fig:2} shows that matter is very weakly coupled
to the scalar field in the solar system ($\alpha_0\approx 0$), and
interferometers like {\sc LIGO} or {\sc VIRGO} might be
insensitive even to large scalar waves. Actually, numerical
calculations\,\cite{novak} suggest that the effects of such
helicity-0 waves are too small to be of observational interest.

There remains nevertheless an indirect way to test for the
presence of a scalar partner to the graviton, in gravitational-wave
observations of inspiralling binaries.\cite{will} Indeed, even if no
helicity-0 wave is detected, the {\it time-evolution\/} of
the detectable helicity-2 chirp depends on the energy flux,
Eq.~(\ref{eq:4}), which can differ significantly from the general
relativistic prediction because of nonperturbative strong-field
effects. If one analyses the data with filters constructed from the
standard general relativistic orbital phase evolution, the
signal-to-noise ratio will thus drop. On the contrary, if one
detects a coalescence using such standard filters, this will
constrain the coupling strength of the bodies to a possible scalar
field. Using such a matched-filter analysis, Will\,\cite{will} has
computed the bounds that a {\sc LIGO} advanced detector could bring
on Brans-Dicke theory (the vertical axis of
Figs.\ref{fig:2}-\ref{fig:3}). We have generalized his results to
the plane $(\alpha_0,\beta_0)$ of generic tensor-scalar
theories.\cite{def9} We found that the detection of a neutron
star-black hole coalescence would exclude all the theories lying
above the hatched straight line of Fig.~\ref{fig:3}. Similarly, the
detection of a double neutron star coalescence would exclude the
bubble of theories labeled NS-NS in this figure. Like binary-pulsar
tests, gravity-wave observations have thus the capability of probing
theories which are strictly indistinguishable from general
relativity in the solar system (namely, those with $\alpha_0=0$ and
$\beta_0$ large and negative). However, we see in
Fig.~\ref{fig:3} that these theories are already excluded by
binary-pulsar data. Paradoxically, this is a good news for the
{\sc LIGO} or {\sc VIRGO} projects: This proves that it is sufficient
to compute the filters within the simpler framework of general
relativity. Of course, this does not reduce the great interest of
such projects. They will provide the first direct observation of
gravitational waves in the wave zone, will (hopefully) lead to
additional confirmations of general relativity through the wave
forms, and will provide important astrophysical information, like
masses and radii of neutron stars, or distance measurements up to
hundreds of Mpc.

\section{Conclusion}

The main conclusion of our study is that strong-field tests of
gravity (binary pulsars and gravitational-wave observations) are
{\it qualitatively\/} different from solar-system experiments. They
constrain theories which are perturbatively equivalent to general
relativity. It is interesting to rewrite in terms of the Eddington
parameters the bound $\beta_0 > -4.5$ that we obtained in Sec.~3. We
find: $(\beta^{\rm PPN} -1)/(\gamma^{\rm PPN} -1) < 1.1$. The singular
($0/0$) nature of this ratio vividly expresses why such a conclusion
could not be obtained in weak-field experiments.

We also underlined that binary pulsars are the best tools to test
alternative theories of gravity. In fact, we have
shown\,\cite{def9} that the best ``scalar probe'' would be a pulsar
orbiting around a black hole companion. One could constrain the
deviations from general relativity at the level $\alpha_0^2
\mathrel{\oalign{$<$\cr$\sim$}} 10^{-6}$, {\it i.e.}, three orders
of magnitude tighter than the present limits.

Our third conclusion is that general relativistic filters can be
used confidently to analyze the data of {\sc LIGO} or {\sc VIRGO}
interferometers. Indeed, even if there exists a scalar partner to
the graviton, binary pulsars already tell us that it is too weakly
coupled to matter to change significantly the chirp templates.

\end{document}